\documentclass[12pt]{article}
\usepackage{fancyhdr}

\usepackage{mathrsfs}
\usepackage[T1]{fontenc}
\usepackage{mathpazo}
\usepackage{setspace}
\usepackage{amsfonts}
\usepackage{amssymb}
\usepackage{amsmath}
\usepackage{epsfig}
\usepackage{latexsym}
\usepackage{color}
\usepackage{graphicx}
\usepackage{nicefrac}
\usepackage[latin1]{inputenc}
\usepackage{pstricks}
\usepackage{slashed}
\usepackage{multirow}
\usepackage{comment}
\usepackage{soul}
\usepackage{hyperref}
\usepackage{cite}

\def\hybrid{\topmargin -20pt    \oddsidemargin 0pt
        \headheight 0pt \headsep 0pt
        \textwidth 6.25in       
        \textheight 9.25in       
        \marginparwidth .875in
        \parskip 5pt plus 1pt   \jot = 1.5ex}

\hybrid

\def\baselinestretch{1.2}

\catcode`\@=11

\def\marginnote#1{}
\newcount\hour
\newcount\minute
\newtoks\amorpm
\hour=\time\divide\hour by60
\minute=\time{\multiply\hour by60 \global\advance\minute by-\hour}
\edef\standardtime{{\ifnum\hour<12 \global\amorpm={am}%
        \else\global\amorpm={pm}\advance\hour by-12 \fi
        \ifnum\hour=0 \hour=12 \fi
        \number\hour:\ifnum\minute<10 0\fi\number\minute\the\amorpm}}
\edef\militarytime{\number\hour:\ifnum\minute<10 0\fi\number\minute}

\def\draftlabel#1{{\@bsphack\if@filesw {\let\thepage\relax
   \xdef\@gtempa{\write\@auxout{\string
      \newlabel{#1}{{\@currentlabel}{\thepage}}}}}\@gtempa
   \if@nobreak \ifvmode\nobreak\fi\fi\fi\@esphack}
        \gdef\@eqnlabel{#1}}
\def\@eqnlabel{}
\def\@vacuum{}
\def\draftmarginnote#1{\marginpar{\raggedright\scriptsize\tt#1}}

\def\draft{\oddsidemargin -.5truein
        \def\@oddfoot{\sl preliminary draft \hfil
        \rm\thepage\hfil\sl\today\quad\militarytime}
        \let\@evenfoot\@oddfoot \overfullrule 3pt
        \let\label=\draftlabel
        \let\marginnote=\draftmarginnote
   \def\@eqnnum{(\theequation)\rlap{\kern\marginparsep\tt\@eqnlabel}%
\global\let\@eqnlabel\@vacuum}  }

\def\preprint{\twocolumn\sloppy\flushbottom\parindent 2em
        \leftmargini 2em\leftmarginv .5em\leftmarginvi .5em
        \oddsidemargin -.5in    \evensidemargin -.5in
        \columnsep .4in \footheight 0pt
        \textwidth 10.in        \topmargin  -.4in
        \headheight 12pt \topskip .4in
        \textheight 6.9in \footskip 0pt
        \def\@oddhead{\thepage\hfil\addtocounter{page}{1}\thepage}
        \let\@evenhead\@oddhead \def\@oddfoot{} \def\@evenfoot{} }

\def\numberbysection{\@addtoreset{equation}{section}
        \def\theequation{\thesection.\arabic{equation}}}

\def\underline#1{\relax\ifmmode\@@underline#1\else
        $\@@underline{\hbox{#1}}$\relax\fi}

\def\titlepage{\@restonecolfalse\if@twocolumn\@restonecoltrue\onecolumn
     \else \newpage \fi \thispagestyle{empty}\c@page\z@
        \def\thefootnote{\fnsymbol{footnote}} }

\def\endtitlepage{\if@restonecol\twocolumn \else \newpage \fi
        \def\thefootnote{\arabic{footnote}}
        \setcounter{footnote}{0}}  

\catcode`@=12
\relax

\def\figcap{\section*{Figure Captions\markboth
        {FIGURECAPTIONS}{FIGURECAPTIONS}}\list
        {Figure \arabic{enumi}:\hfill}{\settowidth\labelwidth{Figure
999:}
        \leftmargin\labelwidth
        \advance\leftmargin\labelsep\usecounter{enumi}}}
 \relax
\def\tablecap{\section*{Table Captions\markboth
        {TABLECAPTIONS}{TABLECAPTIONS}}\list
        {Table \arabic{enumi}:\hfill}{\settowidth\labelwidth{Table
999:}
        \leftmargin\labelwidth
        \advance\leftmargin\labelsep\usecounter{enumi}}}
 \relax
\def\reflist{\section*{References\markboth
        {REFLIST}{REFLIST}}\list
        {[\arabic{enumi}]\hfill}{\settowidth\labelwidth{[999]}
        \leftmargin\labelwidth
        \advance\leftmargin\labelsep\usecounter{enumi}}}
 \relax
\makeatletter
\newcounter{pubctr}
\def\publist{\@ifnextchar[{\@publist}{\@@publist}}
\def\@publist[#1]{\list
        {[\arabic{pubctr}]\hfill}{\settowidth\labelwidth{[999]}
        \leftmargin\labelwidth
        \advance\leftmargin\labelsep
        \@nmbrlisttrue\def\@listctr{pubctr}
        \setcounter{pubctr}{#1}\addtocounter{pubctr}{-1}}}
\def\@@publist{\list
        {[\arabic{pubctr}]\hfill}{\settowidth\labelwidth{[999]}
        \leftmargin\labelwidth
        \advance\leftmargin\labelsep
        \@nmbrlisttrue\def\@listctr{pubctr}}}
 \relax
\makeatother
\newskip\humongous \humongous=0pt plus 1000pt minus 1000pt

\newif\ifdtup

\relax

\def\be{\begin{equation}}
\def\ee{\end{equation}}
\def\ba{\begin{eqnarray}}
\def\ea{\end{eqnarray}}

\def\del{\partial}

\def\a{\alpha}

\def\b{\beta}

\def\g{\gamma}

\def\d{\delta}

\def\m{\mu}

\def\l{\lambda}

\def\s{\sigma}

\def\no{\noindent}

\def\qq{\qquad}

\def\IR{\relax{\rm I\kern-.18em R}}

\def \ha {{1\over 2}}

\def \ov {\over}

\def\IR{\relax{\rm I\kern-.18em R}}
\def\IL{\relax{\rm I\kern-.18em L}}

\def\inv{^{\raise.15ex\hbox{${\scriptscriptstyle -}$}\kern-.05em 1}}

\begin{document}

\renewcommand{\theequation}{\thesection.\arabic{equation}}
\csname @addtoreset\endcsname{equation}{section}

\newcommand{\beq}{\begin{equation}}
\newcommand{\eeq}[1]{\label{#1}\end{equation}}
\newcommand{\ber}{\begin{equation}}
\newcommand{\eer}[1]{\label{#1}\end{equation}}
\newcommand{\eqn}[1]{(\ref{#1})}
\begin{titlepage}
\begin{center}


\vskip0.5in

{\large\bf  All-loop anomalous dimensions in integrable $\lambda$-deformed $\sigma$-models}

\vskip 0.4in

{\bf George Georgiou,$^1$ Konstantinos Sfetsos}$^{2,3}$\ and\ {\bf Konstantinos Siampos}$^{3}$
\vskip 0.1in

\vskip 0.1in
{\em
${}^1$Institute of Nuclear and Particle Physics,\\ National Center for Scientific Research Demokritos,\\
Ag. Paraskevi, GR-15310 Athens, Greece
}
\vskip 0.1in

 {\em
${}^2$Department of Nuclear and Particle Physics,\\
Faculty of Physics, University of Athens,\\
Athens 15784, Greece\\
}
\vskip 0.1in

{\em ${}^3${Albert Einstein Center for Fundamental Physics,\\
Institute for Theoretical Physics, Bern University,\\
Sidlerstrasse 5, CH3012 Bern, Switzerland\\
}}
\vskip 0.1in

{\footnotesize\tt georgiou@inp.demokritos.gr, ksfetsos@phys.uoa.gr, siampos@itp.unibe.ch}


\vskip .5in
\end{center}

\centerline{\bf Abstract}

\no
We calculate the all-loop anomalous dimensions of current operators in
$\lambda$-deformed $\sigma$-models. For the isotropic integrable deformation and for
a semi-simple group $G$ we compute the anomalous dimensions using two different methods.
In the first we use the all-loop effective action and in the second we employ perturbation
theory along with the Callan--Symanzik equation and in conjunction with a duality-type symmetry
shared by these models.
Furthermore, using CFT techniques we compute the all-loop anomalous dimension of bilinear
currents for the isotropic deformation case and a general $G$. Finally we work out 
the anomalous dimension matrix for the cases of anisotropic $SU(2)$
and the two coupling, corresponding to the symmetric coset $G/H$ and a subgroup
$H$, splitting of a group $G$.

\vskip .4in
\noindent
\end{titlepage}
\vfill
\eject

\newpage

\tableofcontents

\noindent

\def\baselinestretch{1.2}
\baselineskip 20 pt
\noindent


\setcounter{equation}{0}
\section{Introduction and motivation}
\renewcommand{\theequation}{\thesection.\arabic{equation}}

Our starting point will be the non-Abelian Thirring model action (for a general 
discussion, see \cite{Dashen:1974gu,Karabali:1988sz}), namely the WZW two-dimensional
conformal field theory (CFT) which will be perturbed by a set of
classically marginal operators that are bilinear in the currents. Namely the total action reads
\be
\label{WZW-pert}
S=S_{{\rm WZW},k}(g) +{1\ov 2\pi} \sum_{a,b=1}^{\dim G} \lambda_{ab} \int \mathrm{d}^2\s\,  J_+^a J_-^b\ ,
\ee
where the WZW action is
\be
S_{{\rm WZW},k}(g) = -{k\ov 4\pi} \int\mathrm{d}^2\sigma\, {\rm Tr}(g^{-1} \del_+ g g^{-1} \del_- g)
+ { k\ov 12\pi} \int_B {\rm Tr}(g^{-1}\mathrm{d}g)^3\ ,
\label{WZW}
\ee
which contains a (left) $\times$ (right) level k representation of an affine Lie algebra G.
The couplings are denoted by $\l_{ab}$.
In what follows we shall need the operator product expansions (OPE) of the currents.
These are better presented for Euclidean world-sheets with complex coordinates $z=\ha(\tau + i \s)$
and $\bar z$.
Adopting a left/right symmetric scheme these read \cite{Witten:1983ar,Knizhnik:1984nr}
\begin{equation}
\begin{split}
\label{OPE}
&J^{a}(z)J^{b}(0)=\frac{\delta_{ab}}{z^2}+\frac{f_{abc}}{\sqrt{k} \ z}\,J^{c}(0)+ {\rm regular}\ ,
\\
&J^{a}(z)\bar{J}^b(0)= {\rm regular} \ .
\end{split}
\end{equation}
In \cite{Kutasov:1989dt} the $\beta$-function for a single (isotropic) coupling $\l$ was computed
to leading order in the $1/k$ expansion but exactly in the $\l$. This was generalized in
\cite{Gerganov:2000mt} for the symmetric couplings. The reader should be aware that
in \cite{Gerganov:2000mt} a different parametrization for the couplings was adopted, see
\cite{Sfetsos:2014jfa} for the notational correspondence.

On the other hand a deformation of the WZW action was introduced by one of the present authors in
 \cite{Sfetsos:2013wia} by
gauging a common symmetry subgroup of the combined action involving the PCM model and the WZW actions. 
Explicitly this action reads
\be
\label{effective.action}
S_{k,\lambda}(g)=\frac{k}{4\pi}\int \mathrm{d}^2\sigma\,J_+DJ_-+
\frac{k}{24\pi}\int_B f_{abc}\,L^a\wedge L^b \wedge L^c+
\frac{k}{2\pi}\int \mathrm{d}^2\sigma\,J_+(\lambda^{-1}-D^T)^{-1}J_-\ ,
\ee
where the first two terms correspond to the WZW model action appropriately rewritten for our purposes. The third term
contains the coupling constants  $\l_{ab}$ assembled as elements of a matrix $\l$. 
{ 
For isotropic couplings the deformation is integrable in the group and symmetric coset cases 
\cite{Sfetsos:2013wia,Itsios:2014vfa,Hollowood:2014rla,Hollowood:2014qma}, which
are embed as solutions of supergravity \cite{Sfetsos:2014cea,Demulder:2015lva}.}
Also we have denoted
\be
\begin{split}
& J^a_+ = -i\, {\rm Tr}(t_a \del_+ g g^{-1}) = R^a_\m \del_+ X^\m \ ,\qq J^a_- = -i\,
{\rm Tr}(t_a g^{-1} \del_- g )= L^a_\m \del_- X^\m\ ,\\
&{ 
R^a:=-i\,{\rm Tr}(t_a\mathrm{d}g\,g^{-1})=R^a_\mu\,\mathrm{d}X^\mu\,,\quad 
L^a:=-i\,{\rm Tr}(t_ag^{-1}\mathrm{d}g)=L^a_\mu\,\mathrm{d}X^\mu\,,}\\
& R^a = D_{ab}L^b \ ,\qq D_{ab}={\rm Tr}(t_a g t_b g^{-1})\ ,
\label{jjd}
\end{split}
\ee
which obey
\be
\mathrm{d}L^a=\frac12\,f_{abc}\,L^b\wedge L^c\,,\qq \mathrm{d}R^a=-\frac12\,f_{abc}\,R^b\wedge R^c\,,\qq
\mathrm{d}D_{ab}=D_{ac}\,f_{cbe}\,L^e\,.
\ee
The matrices $t_a$ obey the commutation relations ${[t_a,t_b]= f_{abc} t_c}$
and are normalized as ${\rm Tr}(t_a t_b)=\d_{ab}$. .

It was conjectured in \cite{Itsios:2014lca,Sfetsos:2014jfa} that \eqn{effective.action} is
the all-loop effective action in $\l$ and to leading order in the $1/k$ expansion
of the non-Abelian Thirring model \eqref{WZW-pert}. This was based on
the following reasoning:
First, for small $\l_{ab}$ the action \eqn{effective.action} becomes the non-Abelian Thirring.
Secondly, both actions share two very important symmetries.
The more obvious one is the invariance under the generalized parity transformation
\be
\label{parity}
\sigma^+\leftrightarrow\sigma^-\,,\qquad g\mapsto g^{-1}\,,\qquad \lambda\mapsto\lambda^T\ .
\ee
 The second and most important for our purposes symmetry is less obvious.
 The following identity  for the action \eqref{effective.action} is obeyed  \cite{Itsios:2014lca,Sfetsos:2014jfa}
\be
\label{sduality}
S_{-k,\lambda^{-1}}(g^{-1})=S_{k,\lambda}(g)\, ,
\ee
which reveals a duality-type symmetry.
This is proved using the transformations
\be
g\mapsto g^{-1}:\qquad D_{ab}\mapsto D_{ba}\,,\qq J_+^a\mapsto-D_{ba}J^b_+\,,\qquad J_-^a\mapsto-D_{ab}J^b_-\,.
\ee
The non-Abelian Thirring action \eqn{WZW-pert} does not have this symmetry at the classical level. However,
using path integral arguments it was proved by Kutasov \cite{Kutasov:1989aw}
that the effective action of the non-Abelian Thirring
model should be invariant
under the above duality-type symmetry $(\l,k)\mapsto (\l^{-1},-k)$ (for $k\gg 1$).

The most compelling reason so far that \eqn{effective.action} is the all-loop effective action in $\l$ for the
non-Abelian Thirring model \eqn{WZW-pert} is the fact that the exact $\b$-function for the couplings $\l_{ab}$
were computed using \eqn{effective.action} in \cite{Itsios:2014lca,Sfetsos:2014jfa} (for symmetric $\l_{ab}$, see also 
\cite{Tseytlin:1993hm}) agree with those obtained with the all-loop summation of the perturbative result using \eqn{WZW-pert}
in \cite{Gerganov:2000mt} (for the case of symmetric $\l_{ab}=\l_{ba}$, for which
results were available in this work). For isotropic $\l_{ab}= \l \,\d_{ab}$ there is also
agreement with the result of \cite{Kutasov:1989dt} as well.

The purpose of this paper is to compute exactly in $\l$ anomalous dimensions of operators using the above effective action.
Our results will be compared with those arising from perturbation theory using CFT techniques, the
Callan--Symanzik equation and the duality-type symmetry \eqn{sduality}. The two approaches are
found in perfect agreement.
Moreover we compute the all-loop anomalous dimension for bilinear currents for the isotropic case, the anisotropic
$SU(2)$ case, as well as  for the case with two { couplings} corresponding to the symmetric coset $G/H$ and a subgroup
$H$ splitting of a group $G$.

\section{All-loop anomalous dimension for isotropic couplings}

In this section we compute the anomalous dimension of the currents $J^a$ for an 
isotropic coupling for a semi-simple group $G$ via the effective action
as well as by using  CFT techniques.

\subsection{Effective action}

\no
On general grounds we have the relation $J^a_{\pm\ {\rm bare}} =Z^{1/2} J^a_\pm$, where
$J^a_{\pm\ {\rm bare}}$ are the bare (unrenormalized) currents, with
$Z$ being the wavefunction renormalization.  To identify this we shall first
consider the group element parametrized as $g= e^{i x^a t_a}$, where $x^a$
will be the coordinates of the deformed $\s$-model action \eqn{effective.action}.
In the limit of small $x^a$'s this effective action \eqref{effective.action} can be written as
\be
\label{bare.eff}
S_{k,\lambda}(g) = 
Z^{-1}\,\int\mathrm{d}^2\sigma\, \del_{+} x^a \del_- x^a + \cdots\,,\quad
Z^{-1}=\frac{k}{4\pi}\,\frac{1+\lambda}{1-\lambda}\,,
\ee
with $J^a_{\pm}=\partial_{\pm}x^a+\cdots$ where the dots refer to higher order interaction terms. Next we note 
that the $x^a$'s are the bare fields as their
diffeomorphisms $\xi_a$ \cite{Itsios:2014lca}
\begin{equation}
\xi_a=-e\,f_{abc}\,\Lambda_{bc}\,,\quad \Lambda=\frac{D-\lambda\mathbb{I}}{\mathbb{I}-\lambda D}\,,\quad
e:=\frac{1}{\sqrt{k(1-\lambda^2)}}\,\frac{\lambda}{1+\lambda}\,,
\end{equation}
vanish in the limit of small $x^a$'s.

\no
The anomalous dimension for the currents is by definition given by
\be
\label{anomalous.current}
\gamma_J:=\mu\frac{\partial\ln Z^{\nicefrac{1}{2}}}{\partial\mu}=\beta\,\frac{\partial \ln Z}{\partial\lambda}\ ,
\ee
where the expression for the $\beta$-function is $\beta(\lambda)=\mathrm{d}\lambda/\mathrm{d}t$,
with $t=\ln{\mu^2}$ and $\m$ the RG energy scale.
Plugging the expression for the $\beta$-function for the case at hand \cite{Kutasov:1989dt,Gerganov:2000mt, Itsios:2014lca}
\begin{equation}
\label{beta-iso}
{
\beta=-\frac{1}{2k}\frac{c_G\,\lambda^2}{(1+\lambda)^2}\leqslant0\ ,
}
\end{equation}
into \eqref{anomalous.current} we find that the all-loop anomalous dimension of the current is given by
 \be
 {
\label{dim.current}
\gamma_J=\frac{c_G\lambda^2}{k(1-\lambda)(1+\lambda)^3}\geqslant0\ ,
}
\ee
where $c_G$ is the second Casimir of the adjoint representation and is related to the structure constants through
${f_{acd}f_{bcd}=-c_G\,\delta_{ab}}$.

\no
Some important comments are in order:

\begin{enumerate}

\item
The above result is invariant under the duality-type symmetry $(\l,k)\mapsto (\l^{-1},-k)$ inherited
by the corresponding invariance of the action \eqref{sduality}.

\item
We prove below that a three-loop in $\l$ perturbative computation results at
\be
{
\label{dim.currentCFT1}
\gamma_J^{FT}= \frac{ c_G}{k}  \left(\lambda^2-2 \lambda^3 + {\cal O}(\lambda^4) \right)\ ,
}
\ee
which is perfect agreement with \eqn{dim.current}. In addition, we shall show that the latter is also
consistent with the Callan--Symanzik equation.

\item
It is interesting to consider the $k\to \infty$ limit in the correlated way \cite{Sfetsos:1994vz,Sfetsos:2013wia}
\be
g = \mathbb{I} + i {v\ov k}\ ,\qq \l= 1-{\kappa^2\ov k}\ ,\qq k\to \infty \ .
\label{gnonabl}
\ee
In that limit the action \eqn{effective.action} becomes \cite{Sfetsos:1994vz}
\be
S = {1\ov 2\pi} \int\mathrm{d}^2\sigma\, \del_+ v^a (\kappa^2 \mathbb{I} + f)^{-1}_{ab} \del_- v^b\ ,\qq f_{ab} = f_{abc} v_c\ ,
\label{sdkjhc}
\ee
which is the well known non-Abelian T-dual of the PCM for a group $G$.
The anomalous dimension \eqn{dim.current} is well defined in that limit becoming
\be
{
\g_J = {c_G\ov 8 \kappa^2}\geqslant0 \ .
\label{fhf}
}
\ee
There is another interesting limit, for $k\to\infty$ and $\lambda\to-1$,  the so called pseudodual limit \cite{Nappi:1979ig},
which we shall study in a subsequent work.

\item
It { was} recently established that the $\l$-deformed models introduced by
\cite{Sfetsos:2013wia} for group and coset spaces and generalized by \cite{Hollowood:2014rla,Hollowood:2014qma}
for semi-symmetric coset spaces, are closely related
\cite{Klimcik:2002zj,Klimcik:2008eq,Vicedo:2015pna,Hoare:2015gda,Sfetsos:2015nya,Klimcik:2015gba}
to the so-called $\eta$-deformed models for group and coset spaces introduced
in \cite{Klimcik:2002zj,Klimcik:2008eq} and in
\cite{Delduc:2013fga,Delduc:2013qra,Arutyunov:2013ega} respectively.
The relation is via Poisson--Lie T-duality
and analytic continuation of coordinates as well as of the parameters of the $\s$-models.
Following the correspondence established in \cite{Hoare:2015gda,Sfetsos:2015nya} we have for the parameters
the analytic continuation
\be
\lambda\mapsto \frac{i-\eta}{i+\eta}\ , \qq  k\mapsto \frac{i}{4t\eta}\ .
\label{analel}
\ee
We easily see that the anomalous dimensions \eqn{dim.current} remain real and become
\be\label{eta-anomal}
{
\gamma_J = \frac{1}{4}\,c_G\ (t\eta)\ \frac{(1+\eta^2)^2}{\eta}\ .
}
\ee
We believe that these are the anomalous dimensions of operators in the $\eta$-deformed
models. They are perturbative in the ($t \eta$)-expansion (like $k$, the product $t\eta$ is a
RG-flow invariant \cite{Sfetsos:2015nya}), but exact in the parameter $\eta$.
Unlike the case of the non-Abelian T-duality limit we considered above it is not
straightforward to establish which operators have these anomalous dimensions.

\end{enumerate}

Note that the above simple-minded background expansion around the identity for the group element
cannot cope with a general matrix $\lambda$, i.e. not proportional to the identity.
In that case the background expansion point will depend non-trivially on $\lambda$.
To isolate the field independent part we have to ascertain the proper vacuum.
An analysis in that spirit was performed recently, again for $\lambda$ proportional to the identity,
in \cite{Appadu:2015nfa} where the expansion of the group element was done around a non-trivial, albeit
associated to two commuting generators, group element.
The choice of these generators has to be arbitrary and this enforces, from
Schur's lemma, $\lambda$ to be proportional to the identity, as it has to
be invariant under a rotation with respect to an arbitrary constant
group element.

\subsection{Field theory}

In this subsection we compute the 2- and 3-loop anomalous dimension of the currents $J^a$ by deriving
\eqn{dim.currentCFT1}, which as mentioned is in perfect agreement with the all-loop expression
\eqref{dim.current} obtained from the effective action \eqref{effective.action}.
Based on this perturbative calculation one can actually argue that under very mild assumptions
the all-loop result has to be given by \eqref{dim.current}.
We shall present this line of { argument} at the end of this section.

In order to find the anomalous dimensions of the currents $J^a$
in the interacting theory one should calculate the correlator of two currents.
On general grounds this correlator will take the form
\be
\label{generic.corr}
G_{ab}(x_1,x_2)=
\langle J^a(x_1)J^b(x_2)\rangle=G_0(\l,k) {\delta_{ab}\ov x_{12}^2}
\left(1+\gamma_J \ln {\frac{\varepsilon^2}
{|x_{12}|^{2}}}\right)+\cdots\, ,
\ee
where $\varepsilon$ is a small distance cut-off regulating the integrals and $x_{ij}=x_i-x_j$. 
{ Notice that the 2-point correlator \eqref{generic.corr} can be rewritten as 
$$G_{ab}(x_1,x_2)=G_0(\l,k) {\delta_{ab}\, \varepsilon^{2\gamma_J} \ov x_{12}^{2(1+\gamma_J/2)} \bar{x}_{12}^{2 \gamma_J/2}}
+\cdots\, . $$ Thus, the engineering
dimensions, holomorphic and anti-holomorhic of our currents are $(1,0)$ and $(0,1)$ respectively
when we are at the conformal point. However, as soon as we turn on interactions the holomorphic dimension becomes
$(1+\gamma_J/2,\gamma_J/2)$. Of course the total anomalous dimension the operator acquires 
is the sum of the two and is $\gamma_J$. Furthermore, notice that the approach  of the previous
section, based on the all-loop effective action provides us directly with the full anomalous dimension $\gamma_J$. 
}

Thus, one has to evaluate the coefficient of the logarithmic term, as well as the finite term $G_0(\lambda)$.
We are seeking the leading and the next-to-leading term in the $1/k$
expansion of the anomalous dimension. The leading term corresponds to the
Abelian part of $G_0(\lambda)$. This is calculated in the appendix \ref{singleG} where we obtain
\be\label{G0}
G_0(\lambda)=\frac{1}{1-\lambda^2}\ .
\ee
This expression will also receive $1/k$ corrections each one of which will be multiplied by a
function of $\l$. We shall show below that the ${\cal O}(1/k)$-correction will be of the
form  $\displaystyle {1\ov k} (\l^3 + {\cal O}(\lambda^4) )$, up to a numerical coefficient which we shall compute below.

We now proceed to identify the logarithmic part of the 2-point function.
We shall need the very basic integrals given by
\begin{equation}\label{id1}
\int  \frac{\mathrm{d}^2z}{(x_1-z)(\bar{z}-\bar{x}_2)}=\pi\ln{|x_{12}|^2}\,,
\end{equation}
\begin{equation}\label{id2}
\int \frac{\mathrm{d}^2z}{(x_1-z)^2(\bar{z}-\bar{x}_2)}=-\frac{\pi}{x_{12}}\ ,
\qq
\int \frac{\mathrm{d}^2z}{( x_1- z)(\bar{z}-\bar{x}_2)^2}=-\frac{\pi}{\bar x_{12}}\,,
\end{equation}
and
\begin{equation}\label{id3}
\int \frac{\mathrm{d}^2z}{(x_1-z)^2(\bar{z}-\bar{x}_2)^2}=\pi^2\delta^{(2)}(x_{12})\,.
\end{equation}
We shall also encounter several times the integral
\begin{equation}
\label{id4}
I(x_1; x_2) = \int \frac{\mathrm{d}^2z}{(z-x_1)(z-x_2)(\bar{z}-\bar{x}_1)}= - {\pi\ov x_{12}}
\ln { \varepsilon^2\ov |x_{12}|^2 }\ .
\end{equation}
This can be computed by writing
$\displaystyle {1\ov (z-x_1)(z-x_2)} = {1\ov x_{12}}\left({1\ov z-x_1} - {1\ov z -x_2}\right)$
and then using \eqn{id1}.

In order to simplify the calculation we shall adopt the regularization scheme of
\cite{Candu:2012xc} in which the integrals are regularized by a short distance cut-off $\varepsilon$.
Due to this the integration domain becomes
\begin{equation}
\label{domain}
D_n=\{(z_1,z_2,\dots ,z_n) \in \mathbb{C}^n: \mid z_i-x_1
\mid > \varepsilon, \mid z_i-x_2\mid > \varepsilon, \mid z_i-z_j \mid  > \varepsilon \}\ ,
\end{equation}
that is all distances { between any} two points cannot be smaller than $\varepsilon$.
In such a case the integral \eqref{id3} is simply replaced by
\begin{equation}
\label{id3mod}
\int_{D_1}  \frac{\mathrm{d}^2z_1}{(x_1-z_1)^2(\bar{z}_1-\bar{x}_2)^2}=0\ ,
\end{equation}
since the $\delta$-function has non-zero support only outside the region of integration.
In the rest of the main part of the paper all of the integrals will be performed in the domain \eqn{domain}.
For notational convenience we shall omit the subscript $D_n$, as in \eqn{id3mod}, from the integration symbol.
Let us point out that the derivation of \eqref{G0} performed in appendix \ref{singleG} is done in the entire complex plane,
since in that case no regularization is required.

Since we shall use CFT techniques, we shall be working with a  Euclidean world-sheet with
complex coordinates $z$ and $\bar z$, i.e. $z=\ha (\tau + i \s)$. Then the perturbation to  the WZW action will be
$\displaystyle {\l\ov \pi} \int \mathrm{d}^2 z\,  J^a \bar J^a$. In the Euclidean regime
in the path integral the action appears as $e^{-S}$. This implies that the factors of $\l$ will contribute as
$\displaystyle {1\ov n!} \left(-\frac{\l}{\pi}\right)^n $ to the $n$-point function.

Obviously the 1-loop contribution to the correlator is vanishing since
$\langle \bar J^{a_1}(\bar z)\rangle =0 $. Hence the first non-vanishing contribution to
the anomalous dimension may come from the 2-loop computation to which we turn now.

\subsubsection*{Calculation at 2-loops}

The 2-loop contribution to the correlator is given by
\be
\label{G2}
G_{ab}^{(2)}={1\ov 2!} \left(\lambda\ov \pi\right)^2\int \prod_{i=1}^2 \mathrm{d}^2z_i
\langle J^a(x_1)J^{a_1}(z_1)J^{a_2}(z_2)J^b(x_2)\rangle
\langle \bar{J}^{a_1}(\bar{z}_1)\bar{J}^{a_2}(\bar{z}_2)\rangle \ .
\ee
All the correlators involved in this integral as well as in similar ones through this paper
will be computed at the CFT point for $\l=0$. Hence
\be
\langle \bar{J}^{a_1}(\bar{z}_1)\bar{J}^{a_2}(\bar{z}_2)\rangle=\frac{\delta_{a_1 a_2}}{\bar{z}_{12}^2}\ .
\label{jjbb}
\ee
The next step is to evaluate the 4-point correlator appearing in \eqref{G2}.
Bearing in mind that we need, within our approximation, at most two factors involving
the non-Abelian part of the current-current OPE we obtain
\begin{equation}
\label{4-hol}
\begin{split}
&\langle J^a(x_1)J^{a_1}(z_1)J^{a_2}(z_2)J^b(x_2)\rangle
= {1\ov \sqrt{k}} {f_{ a_1 a c}\ov z_1-x_1}\langle J^c(x_1)J^{a_2}(z_2)J^b(x_2)\rangle
\\
&\phantom{}
 +  {1\ov \sqrt{k}} {f_{ a_2 a c}\ov  z_2-x_1} \langle J^c(x_1)J^{a_1}(z_1)J^b(x_2)\rangle+
{1\ov \sqrt{k}} {f_{ a b c}\ov  x_{12}}\langle J^c(x_2)J^{a_1}(z_1)J^{a_2}(z_2)\rangle + \dots \ .
\end{split}
\end{equation}
Before we proceed with the evaluation of the above three terms we point out that
we have not displayed terms of order one coming from the double pole in \eqn{OPE} as they
correspond to either bubble diagrams or they will provide contributions proportional to
the vanishing, in our regularization scheme, integral \eqn{id3mod}.
In general, the 3-point function for the currents is\footnote{Notice that this expression is fully consistent with the 3-point function of primary
operators \cite{Georgiou:2010an}.}
\be
\langle J^{a_1}(x_1)J^{a_2}(x_2)J^{a_3}(x_3)\rangle = {1\ov \sqrt{k}}
{f_{a_1 a_2 a_3}\ov x_{12}x_{13}x_{23}}\ .
\label{kjhk3}
\ee
Bearing in mind that $a_1= a_2$ due to \eqn{jjbb} we see that the last term in \eqn{4-hol}
does not contribute to the integral in \eqn{G2}. In addition, we easily see that the first two terms in \eqn{4-hol}
contribute equally. Hence
\be
G_{ab}^{(2)} =- {\l^2 \ov \pi^2 k } {c_G \d^{ab}\ov x_{12}}
\int {\mathrm{d}^2z_1 \mathrm{d}^2 z_2 \ov (z_1-x_1) (x_1-z_2) (z_2-x_2)\bar z_{12}^2}
 =- {\l^2 \ov \pi k } {c_G \d^{ab}\ov x_{12}} I(x_1; x_2) \  ,
\ee
where in order to perform the integral over $z_1$ we have used \eqn{id2} and the remaining $z_2$
integration gave the integral defined in \eqn{id4}. Hence, we eventually find that the 2-loop contribution to the correlator
\begin{equation}
\label{G2.semi}
G_{ab}^{(2)}=\frac{\lambda^2}{k}\frac{c_G \delta_{ab}}{x_{12}^2}\ln{\frac{\varepsilon^2} {|x_{12}|^{2}}}\ .
\end{equation}

\subsubsection*{Calculation at 3-loops }

The 3-loop contribution to the two-point function takes the form
\begin{equation}
\label{G3}
G_{ab}^{(3)}=-{1\ov 3!} \left(\lambda\ov \pi\right)^3
\int \prod_{i=1}^3 \mathrm{d}^2z_i \langle J^a(x_1)J^{a_1}(z_1)J^{a_2}(z_2)J^{a_3}(z_3)J^b(x_2)\rangle
\langle \bar{J}^{a_1}(\bar{z}_1)\bar{J}^{a_2}(\bar{z}_2)\bar{J}^{a_3}(\bar{z}_3)\rangle\ .
\end{equation}
The 3-point barred correlator is given by the analogue of \eqref{jjbb} for anti-holomorhic currents.

Subsequently, one has to make the contractions of the unbarred $J$'s.
Taking into account that the current $J^{a}$
can be contracted with $J^{a_1}$ through a $\delta$- or through an $f$- term we obtain the following expression
for the 5-point correlator
\begin{equation}\label{a1acon}
\begin{split}
&\langle J^a(x_1)J^{a_1}(z_1)J^{a_2}(z_2)J^{a_3}(z_3)J^b(x_2)\rangle=
{1\ov \sqrt{k}}
\frac{\delta_{a_1 a}}{(x_1-z_1)^2}\frac{f_{a_2 a_3 b}}{z_{23}(z_2-x_2)(z_3-x_2)}
\\
&+{1\ov \sqrt{k}} \frac{f_{a a_1 c}}{x_1-z_1}\left(\frac{\delta_{a_2c}\delta_{a_3 b}}{z_{12}^2(z_3-x_2)^2}
+ \frac{\delta_{c a_3}\delta_{a_2 b}}{z_{13}^2(z_2-x_2)^2}\right) +  \cdots\   .
\end{split}
\end{equation}
In the above we have omitted several terms denoted by the dots.
Firstly, since there is a factor of $1/\sqrt{k} f_{a_1 a_2 a_3}$ coming from the anti-holomorphic 3-point function
we have not written terms with two $f$'s in the 4-point function in the second line above since that would
be subleading in the $1/k$ expansion. In addition, we have omitted a possible third term in
the second line of \eqref{a1acon} proportional to $\delta_{a_2 a_3}\delta_{b c}$.
This term  will not contribute to the integral when multiplied by the factor of $f_{a_1 a_2 a_3}$ originating from the
anti-holomorphic currents.
The dots stand also for two terms involving the singular part of the OPE of the current $J^a$
with $J^{a_2}$ and $J^{a_3}$.
Because our integral is invariant under the permutation of the interaction vertices each of these
two terms will give a contribution which is identical to the explicitly written contribution
coming from the contraction of the $J^{a}$ and $J^{a_1}$ currents above.
Finally, the contraction involving the two external currents will be zero
since when this is through a $\delta$-term it will correspond
to a bubble diagram and should be, thus, omitted, while when it is through an $f$-term
it will result into a 4-point correlator proportional either
to $\delta_{a_1 a_2}$, or $\delta_{a_2 a_3}$ or $\delta_{a_1 a_3}$ which when multiplied by
$f_{a_1 a_2 a_3}$, coming from the 3-point function of the anti-holomorphic currents,  will give zero.

We now proceed to evaluate the corresponding triple integrals.
The first line of \eqref{a1acon} results into
\begin{equation}\label{deltainit}
G_{1,ab}^{(3)}=  \frac{\lambda^3}{2 \pi^3}\frac{c_G \delta_{a b}}{k}\int
\frac{\mathrm{d}^2z_ 1 \mathrm{d}^2z_ 2  \mathrm{d}^2z_ 3 }
{\bar{z}_{12}\bar{z}_{23}\bar{z}_{13}z_{23}(x_1-z_1)^2(z_2-x_2)(z_3-x_2)}.
\end{equation}
Using the identity
$\displaystyle {1\ov \bar{z}_{12}\bar{z}_{13}}
= {1\ov \bar{z}_{23}}\left({1\ov \bar{z}_{12}} - {1\ov \bar{z}_{13}}\right)$
we split the last integral into two. Subsequently, we perform the $z_1$ integral and
notice that the two integrals which result are equal to each other under the exchange
$z_2 \longleftrightarrow z_3$.
Then we rewrite $\displaystyle {1\ov z_{23}(z_3-x_2)} = {1\ov z_2-x_2}\left({1\ov z_{23}} + {1\ov z_3-x_2}\right)$
to obtain
\begin{equation}\label{deltaintermed}
\begin{split}
&G_{1,ab}^{(3)}=  \frac{\lambda^3}{\pi^2}\frac{c_G\delta_{a b}}{k}\int \mathrm{d}^2z_2 \mathrm{d}^2z_3
\Big(\frac{1}{z_{23}\bar{z}_{23}^2(z_2-x_1)(z_2-x_2)^2}
\\
&\phantom{xxxxxx} + \frac{1}{\bar{z}_{23}^2(z_2-x_1)(z_2-x_2)^2(z_3-x_2)}\Big ).
\end{split}
\end{equation}
The first integral of \eqref{deltaintermed} is zero,\footnote{
The zero comes from the $z_3$ integral.
This $z_3$ integration can be written as
\begin{equation*}
\int \mathrm{d}^2z_3\, \frac{1}{z_{23}\bar{z}_{23}^2}=-\int \mathrm{d}^2z\, z\, \frac{1}{\mid z\mid ^4}=0, \qquad z=z_{23}.
\end{equation*}
In order to see that the last integral is zero just write $z=x + i y$. Then the integrand is odd in $x$ and $y$ and is thus zero when
$x$ and $y$ are integrated from $-\infty$ to $\infty$.
} while the second one will give after performing the $z_3$ integration
\be
\begin{split}
\label{deltafinal}
& G_{1,ab}^{(3)}=   \frac{\lambda^3}{\pi}\frac{c_G\delta_{a b}}{k}\int
\frac{\mathrm{d}^2z_2}{(\bar{x}_2-\bar{z}_2)(z_2-x_1)(z_2-x_2)^2}
=-\frac{\lambda^3}{\pi}\frac{c_G\delta_{a b}}{k} \del_{x_2}  I(x_2; x_1)
\\
& \phantom{xxxx}
= -{\lambda^3\ov k} \frac{c_G \delta_{a b}}{ x_{12}^2}\ln{e\,  \varepsilon^2 \ov |x_{12}|^2}\ ,
\end{split}
\ee
where we note the presence of $e$, i.e. the basis of the natural logarithm. This will contribute to  the
overall normalization of the 3-point function \eqn{generic.corr}.

Next we turn to the second line of \eqref{a1acon}.
The two terms in it give equal contributions to the integral in \eqn{G3}.
This is the case because the triple integral is symmetric under the
 relabelling of $a_2 \longleftrightarrow a_3$
and  $z_2 \longleftrightarrow z_3$.
Thus the result reads
\begin{equation}\label{f-init}
G_{2,ab}^{(3)}= \frac{\lambda^3}{\pi^3}\frac{ c_G\delta_{a b}}{k}\int
\frac{\mathrm{d}^2z_1 \mathrm{d}^2z_2 \mathrm{d}^2z_3 }
{\bar{z}_{12}\bar{z}_{23}\bar{z}_{13}z_{12}^2(x_1-z_1)(z_3-x_2)^2}\ .
\end{equation}
By using the identity $\displaystyle {1\ov \bar{z}_{12}\bar{z}_{23}}
= {1\ov \bar{z}_{13}}\left({1\ov \bar{z}_{12}} + {1\ov \bar{z}_{23}}\right)$ one obtains
two integrals.
The first contributes zero since the $z_3$ integration gives $\delta^{(2)}(x_2-z_1)$
which vanishes inside our integration domain.
To evaluate the second integral we first perform the $z_2$ integral and obtain
\begin{equation}\label{f-intermed}
\begin{split}
G_{2,ab}^{(3)}=  \frac{\lambda^3}{\pi^2}\frac{c_G\delta_{a b}}{k}\int
\frac{\mathrm{d}^2z_1 \mathrm{d}^2z_3}{z_{31}\bar z_{13}^2(x_1-z_1)(z_3-x_2)^2}\ .
\end{split}
\end{equation}
Subsequently, we employ the identity
$\displaystyle {1\ov z_{31}(x_1-z_1)} = {1\ov x_1-z_3}\left({1\ov z_{31}} - {1\ov x_1-z_1}\right)$
to rewrite it as the difference of two integrals.
The first integral will be zero once we perform the $z_1$ integration.
After performing the $z_1$ integration, the second integral becomes
\begin{equation}\label{f-intermed2}
\begin{split}
&G_{2,ab}^{(3)}= \frac{\lambda^3}{\pi}\frac{c_G\delta_{a b}}{k}\int
\frac{\mathrm{d}^2z_3}{(\bar{x}_1-\bar{z}_3)(x_1-z_3)(z_3-x_2)^2}
\\
&\phantom{xxxx} =  \frac{\lambda^3}{\pi}\frac{c_G\delta_{a b}}{k} \del_{x_2} I(x_1; x_2)
\end{split}
\end{equation}
Using \eqn{id4} we get the final result
\begin{equation}\label{f-final}
\begin{split}
&G_{2,ab}^{(3)}= -\frac{\lambda^3}{k}\frac{c_G \delta_{a b}}{x_{12}^2}\ln{e\, \varepsilon^2 \ov |x_{12}|^2}\ ,
\end{split}
\end{equation}
which is the same expression as that in \eqn{deltafinal}.

\no
Putting together the results of the 2-loop and 3-loop
calculations \eqref{G2.semi}, \eqref{deltafinal} and \eqref{f-final},  we obtain the final result
for the 2-point function which up to 3-loops reads
\begin{equation}
G_{ab}(x_1,x_2)
=\frac{\delta_{ab}}{x_{12}^2}\Big(1  +{ \lambda^2}- 2 {c_G\ov k}\l^3
+{ c_G\ov k} (\lambda^2-2 \lambda^3)
\ln{\frac{\varepsilon^2}{|x_{12}|^2}} + {\cal O}(\lambda^4)\Big).
\end{equation}
Comparing with the general form of the 2-point function \eqn{generic.corr}
we can now straightforwardly read off the anomalous dimension of the currents $J^a$ and $\bar{J}^a$ and the
$1/k$ correction to the overall normalization. The anomalous dimension is
given by \eqn{dim.currentCFT1} mentioned at the beginning of this section.
For the wavefunction normalization it is easily found that
\be
G_0(\l,k) = {1\ov 1-\l^2} -{1\ov k}\left(2 c_G \l^3 + {\cal O}(\lambda^4) \right)  + {\cal O}\left(1\ov k^2\right)\ .
\ee

\subsubsection*{All-loop expression}
One can actually use the perturbative result \eqref{dim.currentCFT1}, in fact the 2-loop result will be enough,
to argue that the all-loop expression for the anomalous dimension should be given by \eqref{dim.current}.
Recall the symmetry of the effective action \eqn{sduality}
argued also in \cite{Kutasov:1989aw} from the path integral view point for the non-Abelian Thirring model.
 Thus the anomalous dimensions we are after should be invariant under the simultaneous transformation
of the two couplings $\lambda\rightarrow 1/\lambda$ and $k\rightarrow -k$ for large values of $k$.

Motivated by the expression for the wavefunction renormalization \eqn{zeff} that followed
from the effective action \eqref{effective.action} we shall make the mild assumption
that the only points in the complex $\l$-plane where the anomalous dimension can have poles are at $\lambda=\pm 1 $.
This assumption can also be further justified by the fact that,
as discussed in \cite{Kutasov:1989dt}, the 2-point correlators
can be expressed as covariant derivatives of the effective potential calculated in the same work.
The only points where the effective potential has poles are at $\lambda=\pm 1$.
This implies that the same holds true for the 2-point functions.
As a consequence the anomalous dimension should acquire the following form
\begin{equation}
\label{ass.form}
\gamma_J^{FT}=\frac{ c_G  \lambda^2 f(\lambda)}{k(1-\lambda)^m(1+\lambda)^n}\ ,
\end{equation}
where the function $f(\lambda)$ should be everywhere analytic and also $f(0)=1$
so that \eqref{ass.form} agrees with the leading term in \eqref{dim.currentCFT1}.
We shall now argue that the form of the Callan--Symanzik equation for the 2-point function $G_2$
\begin{equation}\label{Callan}
\mu \frac{\partial G_2}{\partial \mu}+\beta(\lambda)\frac{\partial G_2}{\partial \lambda}
+2 \gamma_J(\lambda)G_2=0 \ .
\end{equation}
implies that $m=1$ and $n=3$.
On general grounds, the expression for the all-loop 2-point function will acquire the following form
\begin{equation}
\label{2-pointgen}
G_2=\frac{g(\lambda,\mu^2x_{12}^2,\gamma_J(\lambda))}{(1+\lambda)^a (1-\lambda)^b}
{1\ov x_{12}^2}\ .
\end{equation}
Notice that we have factorised the poles of the 2-point function at $\lambda=\pm 1$
with some exponents $a$ and $b$.
The numerator $g(\lambda,\mu^2x_{12}^2,\gamma_J(\lambda))$ depends on $\lambda$
both explicitly, as well as implicitly through the anomalous dimension $\gamma_J$.
Plugging \eqref{2-pointgen} into  \eqref{Callan} we obtain
\begin{equation}
\label{subst}
\mu \frac{\partial g}{\partial \mu}
+ \beta \frac{\partial g}{\partial \lambda}
+ \beta \frac{\partial \gamma_J}{\partial \lambda}\frac{\partial g}{\partial \gamma_J}
+\beta g \frac{b-a+(b+a)\lambda}{(1+\lambda) (1-\lambda)}
+ 2\gamma_J g=0\ , \end{equation}
where we have omitted a common overall factor corresponding to the inverse of  $(1+\lambda)^a (1-\lambda)^b x_{12}^2$.
Since $\b$ and $\g_J$ are of order $1/k$, all terms in the above expression are of order $1/k$ except for
the third term which is of order $1/k^2$ and hence it
can be safely ignored.
For generic values of $a$ and $b$ the leading behaviour of \eqref{subst} close to the $\lambda=\pm 1$
poles come solely from the last two terms of  \eqref{subst}, that is all other terms of \eqref{subst}
can be ignored. Then the only way to satisfy \eqref{subst}
is to demand that $\gamma_J \sim \frac{1}{(1-\lambda)(1+\lambda)^{3} }$. This is so because
the $\beta$-function behaves as $\beta \sim \frac{1}{(1+\lambda)^{2}}$ around $\lambda=- 1$ and is
analytic around $\l=1$.
In conclusion we have found that $m=1$ and $n=3$.

Subsequently, we employ invariance of $\gamma_J$ under the aforementioned duality-type transformation of the couplings.
This leads to a constraint on the function $f$ appearing in \eqref{ass.form}, namely that
$f(1/\lambda)=f(\lambda)$. Taking into account  that $f(\lambda)$ is analytic in the complex $\l$-plane and that
$f(0)=1$ we have that $f(\infty)=1$. Hence, according to Cauchy--Liouville's theorem,
that every bounded entire function must be constant, we conclude that $f(\l)=1$.

We conclude that the expression for the all-loop anomalous dimension that is consistent with our perturbative
result and with the symmetries of the theory is
given by \eqref{dim.current}.\footnote{
This result assumes that $b-a+(b+a)\lambda\neq1\pm\l$. However, if $a=0$ and $b=1$ we may have  the plus sign.
Then, it turns out that in that case
\begin{equation*}
{
\gamma_J=\frac{c_G\,\lambda^2 f(\lambda)}{k\,(1-\lambda)(1+\lambda)^2}\ .
}
\end{equation*}
Imposing invariance under the duality-type transformation for the couplings we find that $f(1/\l)= \l f(\l)$,
which since $f(0)=1$ implies that $f(\infty)=0$. This is is not the same constant as its should be
according to  Liouville's theorem. Hence, this expression for the anomalous dimension should be rejected.
With similar considerations we may disregard the minus sign.
}

\section{Anomalous dimension of the bilinear current}
\label{bilinear}

In this section we derive the all-loop in $ \lambda$ anomalous dimension matrix of
the composite current-bilinear operator perturbing the WZW model in \eqn{WZW-pert}. For notational
convenience let us denote this set of operators by ${\cal O}_i$ and the couplings by $\l^i$.
In that way we shall avoid a double index notation and we shall keep the discussion general.

Knowing the anomalous dimension will allow us to determine whether the
classically marginal operators ${\cal O}_i$ become relevant or irrelevant as soon as the deformations $\lambda^i$ are turned on.
Furthermore, it might give us insights into the behaviour of the theory as it flows towards the IR towards the
point $\lambda=1$.
To this end, one should evaluate the 2-point correlation functions
$g_{ij}=\langle{\cal O}_i(x_1) {\cal O}_j(x_2) \rangle |x_{12}|^4$, i.e. the Zamolodchikov metric.
As discussed in \cite{Kutasov:1989dt} the 2-point function ${  G_{ij}=\langle {\cal O}_i{\cal O}_j\rangle}$ and the 
Zamolodchikov metric { $g_{ij}$}
 take the following form
\begin{equation}\label{2-point}
\begin{split}
&{ G_{ij}\sim G_{0|ij}(\lambda^i,k)}\,|x_{12}|^{-2(2+\gamma)}={ G_{0|ij}(\lambda^i,k)}
|x_{12}|^{-4}\left(1+\gamma \ln {\frac{\varepsilon^2}
{|x_{12}|^2}}+\dots\right)\,,\\
&g_{ij}(\lambda^i,t)=g_{ij}^{(0)}(\lambda^i)+2\pi t \nabla_i\nabla_j V+ {\cal O}(t^2)=g_{ij}^{(0)}(\lambda^i)+
t \nabla_{(i}\beta_{j)}+ {\cal O}(t^2)\,,
\end{split}
\end{equation}
where $t=\ln\left({|x_{12}|^{2}\mu^2}\right)$, $\pi\nabla_iV=\beta_i$ and $A_{(ij)}:=A_{ij}+A_{ji}$.
In \eqn{2-point} $\mu$ is an arbitrary renormalization scale, $g_{ij}^{(0)}$ is the finite part of the 2-point correlator
and $V$ is the effective potential. Also, the connections appearing
in the covariant derivatives are calculated with respect to the metric $g_{ij}^{(0)}$ and $\partial_i=\frac{\partial}{\partial\l^i}$.
Thus, we see that the 2-point functions have the geometrical interpretation of a metric
in the curved space whose coordinates are the couplings of the theory ${\lambda^i}$.
We should also mention that we raise and lower the indices of the $\beta$-functions
$\beta^i(\lambda^i)=\frac{\mathrm{d} \lambda^i}{\mathrm{d} t}$
with the metric $g^{(0)}_{ij}$, that is $\beta_i=g^{(0)}_{ij}\beta^j$.

From \eqn{2-point} one can straightforwardly read the anomalous dimension matrix $\gamma_i{}^j$.
It is given by the coefficient of the logarithmic term
after one factorizes the finite part $g_{ij}^{(0)}$. In conclusion we get
\begin{equation}
\label{gamma}
\begin{split}
&\gamma_i{}^j(\lambda^i)=-\nabla_{(i}\beta_{k)}g^{(0)kj}=-\nabla_{i}\beta^{j}-\nabla^{j}\beta_{i}
\\
&\phantom{xxxx} =-\partial_i\beta^j-g^{(0)jm}\left(g^{(0)}_{in}\partial_m\beta^n+\beta^n \partial_n g^{(0)}_{im}\right)\,,\
\end{split}
\end{equation}
transforming as a mixed tensor under diffeomorphisms of $\lambda^i$.
Hence the anomalous dimension matrix for the closed set of operators ${\cal O}_i$ is given
by the covariant derivatives of the corresponding $\beta$-functions with respect to the couplings $\lambda^i$.
For the theories at hand \eqn{WZW-pert}, the leading in $k$ term of the beta-function is proportional to $1/k$
and are known to all-orders in the coupling constants $\lambda^i$.
Since our aim is to calculate the leading in $k$ anomalous dimension matrix it is enough to know the part of
$g_{ij}^{(0)}$ which is independent of $k$, but to all orders in the couplings $\lambda^i$.
This fact amounts to performing all contractions of the currents by completely ignoring the term proportional
 to $f_{abc}$ in the first equation of \eqn{OPE} and integrating over the entire complex plane, since there is 
 no need for regularization.

\subsection{Isotropic case}

As an important example, we shall apply the general formalism for
the calculation of the all-loop anomalous dimension when the perturbing operator
is $J^{a} \bar{J}^{a}$, that is for the case of the isotropic non-Abelian Thirring model.
In such a case the matrix $\gamma_i{}^j$
becomes a single (scalar) function of $\lambda$. 
Using  \eqref{gamma} we easily find
\begin{equation}
\begin{split}
\gamma = -2 \beta'(\lambda)- \beta \frac{g^{(0)'}_{11}}{g^{(0)}_{11}}\ ,
\end{split}
\end{equation}
where  the prime denotes differentiation with respect to $\lambda$.
Using the beta-function for the isotropic non-Abelian Thirring model \eqref{beta-iso}
as well as the $k$-independent part of the metric
\begin{equation}\label{G-iso}
g^{(0)}_{11}=\frac{\dim G }{(1-\lambda^2)^2}\ ,
\end{equation}
proven in appendix \ref{compositeG}, we easily obtain that
\be
\gamma =
{2 c_G\ov k} \frac{ \lambda (1-\lambda(1-\lambda))}{(1-\lambda)(1+\lambda)^3}\geqslant0\ .
\label{jhjf}
\ee
Unlike the anomalous dimension for the currents which is of ${\cal O}(\l^2)$, for
the composite current-bilinear operator the first correction is of ${\cal O}(\l)$.
The expression \eqn{jhjf} is indeed
invariant under the duality-type transformation $\lambda\rightarrow 1/\lambda,\, k \rightarrow -k$.
In addition, in the correlated $k\to \infty$ limit \eqn{gnonabl} we find $\displaystyle \g\to {4 c_G\ov \kappa^2}$. This should be the anomalous dimension of the bilinear $\del_+ v\del_- v$ in the $1/\kappa^2$ expansion for the
non-Abelian T-dual action \eqn{sdkjhc}.

Finally, we include the leading 1-loop contribution to the anomalous dimension.
We have that {  the contribution to the two-point function \eqref{2-point} reads}
\be
\begin{split}
& { G^{(1)}} = - {\l\ov \pi} \int \mathrm{d}^2 z\, \langle J^a(x_1) J^c (z) J^b (x_2)\rangle
\langle \bar J^a(\bar x_1) \bar J^c (\bar z) \bar J^b (\bar x_2)\rangle
\\
& \phantom{xxx}
 = - {{c_G}\ov k} {\l\ov \pi} {\dim G\ov |x_{12}|^2} \int {\mathrm{d}^2 z \ov (x_1 -z)(x_2-z) (\bar x_1 -\bar z)(\bar x_2-\bar z)
 }\ ,
 \end{split}
\ee
where we have used \eqn{kjhk3}. To compute the integral we write
\begin{equation*}
{1 \ov (x_1 -z)(x_2-z) (\bar x_1 -\bar z)(\bar x_2-\bar z)  } = {1\ov |x_{12}|^2}
\left( {1\ov x_1 - z} -  {1\ov x_2 - z}\right)  \left( {1\ov \bar x_1 - \bar z} -  {1\ov \bar x_2 - \bar  z}\right)  \
\end{equation*}
and then use \eqn{id1}. We finally find that
\be
{ G^{(1)}} =  {{ 2}\ov k} {c_G  \dim G\, \lambda\ov |x_{12}|^4} \ln  {\varepsilon^2\ov |x_{12}|^2 }\ .
\ee
Comparing with \eqn{2-point} we find that $\displaystyle \g ={2c_G\ov k}  \l $, which is
consistent with \eqn{jhjf} to that order in the  small $\l$ expansion.

\subsection{Beyond the isotropic case}

In this section we shall generalize the treatment of previous subsection for cases with anisotropy.
We are going to derive the anomalous dimension matrix for the cases of the: i) anisotropic non-abelian
Thirring model with group  $G=SU(2)$ and ii) two coupling corresponding to the symmetric coset $G/H$ and a subgroup
$H$ splitting of a group $G$.

\subsection*{The $SU(2)$ case}

We shall assume that the three operators ${\cal O}_i, \,\,\,i=1,2,3$
which perturb the CFT enter in the combination $\sum\limits_i\lambda^i/\pi\,{\cal O}_i$,
where ${\cal O}_i(z,\bar{z})=J^{i}(z)\bar{J}^{i}(\bar{z})\,\,\,i=1,2,3$ (no summation).
The crucial observation is the leading in $k$ part of the 2-point correlator $g^{(0)}_{ij}$ is a diagonal matrix with the following entries
\begin{equation}\label{Gij-anis}
g^{(0)}_{ij}=\frac{\delta_{ij}}{(1-(\lambda^i)^2)^2}\,,
\end{equation}
generalizing the result of the appendix \ref{compositeG}.
All non-diagonal elements are zero because it is impossible to have any connected diagram contributing to the 2-point function
$\langle {\cal O}_i {\cal O}_j \rangle$ when $i\neq j$. This is no longer true when the non-abelian terms proportional to $f_{abc}$ in the current OPE
are taken into account when performing the contractions of the currents. But these terms bring additional suppression  in powers of $1/\sqrt{k}$.
Since the beta functions are already proportional to
$1/k$ and since we are after the leading in $k$ anomalous dimension matrix we can safely ignore all non-diagonal elements of $g^{(0)}_{ij}$.
Using the same line of argument, it is not difficult { to convince oneself} that the calculation of the diagonal terms boils down to the one
we have performed in the case of the isotropic non-abelian Thirring model. This is so because the only interaction vertex that can contribute to the
$\langle{\cal O}_i {\cal O}_i \rangle$ correlator, in the leading in $k$ expansion, is the one which involves the operator ${\cal O}_i$,
that is $\frac{\lambda^i}{\pi}{\cal O}_i(z,\bar{z})$. The other two interaction vertices will have to combine into bubbles since
in the abelian approximation
one can not contract currents with different group indices. Consequently, the 2-point correlator will take the form of \eqn{Gij-anis}. We are now in position to write down the final expression for the $3\times 3$ anomalous dimension matrix.
Using \eqn{gamma} we easily find
\begin{equation}
\begin{split}
\label{gamma-anis}
&\gamma_i{}^i=-2 \partial_i \beta^i-\frac{1}{g^{(0)}_{ii}}\partial_i g_{ii}^{(0)}\beta^i\,,\quad \rm{no\,\, summation \,\,in \,\,i}=1,2,3\,, \\
&\gamma_i{}^j=-\partial_i \beta^j-\frac{(1-(\lambda^j)^2)^2}{(1-(\lambda^i)^2)^2}\partial_j \beta^i,\,\,\,\,\,\,i\neq j\,.
\end{split}
\end{equation}
Plugging in \eqn{gamma-anis} the values of the exact in $\lambda$ beta functions given by \cite{Gerganov:2000mt}
\begin{equation}\label{beta-anis}
\beta^1=-\frac{2}{k}\frac{(\lambda^2-\lambda^3\lambda^1)(\lambda^3-\lambda^1
\lambda^2)}{(1-(\lambda^2)^2)(1-(\lambda^3)^2)},\quad \text{cyclic in 1,2 and 3}\,,
\end{equation}
we obtain the all-loop expression for the anomalous dimension matrix
\be
\label{anomalous.su2}
\begin{split}
&\gamma_1{}^1=\frac4k\,\frac{4\l^1\l^2\l^3-(1+(\l^1)^2)((\l^2)^2+(\l^3)^2)}{(1-(\l^1)^2)(1-(\l^2)^2)(1-(\l^3)^2)}\,,\\
&\gamma_1{}^2=\frac4k\,\frac{(1+(\l^1)^2)(1+(\l^2)^2)\l^3-2\l^1\l^2(1+(\l^3)^2)}{(1-(\l^1)^2)^2(1-(\l^3)^2)}\,,\\
&\gamma_2{}^1=\frac4k\,\frac{(1+(\l^1)^2)(1+(\l^2)^2)\l^3-2\l^1\l^2(1+(\l^3)^2)}{(1-(\l^2)^2)^2(1-(\l^3)^2)}\,,
\end{split}
\ee
and cyclic in $1,2,3.$ Note that the matrix \eqn{anomalous.su2} transforms as a mixed tensor
under the diffeomorphism endowed by \eqref{sduality}
\be
\label{symmetry}
\gamma_i{}^j((\lambda^i)^{-1},-k)=\left(\frac{\l^i}{\l^j}\right)^2\gamma_i{}^j(\lambda^i,k)\,.
\ee

\subsection*{The two coupling case using a symmetric coset}

Let's split the group index into a part corresponding to a subgroup $H$ of $G$ and the
rest { belonging} to the coset $G/H$. We shall keep denoting by Latin letters the subgroup indices and by
Greek letters the coset indices.
Consider the case in which the matrix $\l_{AB}$ has elements
\be
\l_{ab} = \l^1 \d_{ab}\ ,\qq \l_{\a\b} = \l^2 \d_{\a\b}\ .
\ee
It turns out that the above restriction is consistent only for symmetric coset spaces $G/H$
\be
\begin{split}
\label{RG.sub.coset}
&\frac{\mathrm{d} \l^1}{\mathrm{d} t}=-\frac{c_G(\l^2)^2(1-(\l^1)^2)^2+c_H((\l^1)^2-(\l^2)^2)
(1-(\l^1)^2(\l^2)^2)}{2k(1+\l^1)^2(1-(\l^2)^2)^2} \,,\\
&\frac{\mathrm{d} \l^2}{\mathrm{d} t}=-\frac{c_G\l^2(\l^1-(\l^2)^2)}{2k(1+\l^1)(1-(\l^2)^2)} \,.
\end{split}
\ee
The crucial observation is the leading in $k$ part of the 2-point correlator $g^{(0)}_{ij}$ ($i=1,2$) is a diagonal $2\times2$ matrix with the following entries
\begin{equation}\label{Gij-coset}
g^{(0)}_{11}=\frac{\mathrm{d}_H}{(1-(\lambda^1)^2)^2}\,,\qq g^{(0)}_{22}=\frac{\mathrm{d}_G-\mathrm{d}_H}{(1-(\lambda^2)^2)^2}\,,
\end{equation}
where $1$ and $2$ correspond to the subgroup and coset respectively; $\mathrm{d}_G:=\dim G$, $\mathrm{d}_H:=\dim H$.
Plugging in \eqn{gamma} the above expressions for the metric and the flow, we find the anomalous
dimension matrix
\begin{equation*}
\label{anomalous.coset}
\begin{split}
&\gamma_1{}^1=\frac2k\frac{c_H\l^1(1-(1-\l^1)\l^1)(1+(\l^2)^4)+(\l^2)^2(c_H(1-2\l^1-2(\l^1)^3+(\l^1)^4)-c_G(1-(\l^1)^2)^2)}
{(1-\l^1)(1+\l^1)^3(1-(\l^2)^2)^2}\,,\\
&\gamma_1{}^2=\frac{1}{2k}\frac{(c_G(\mathrm{d}_G+\mathrm{d}_H)-2c_H \mathrm{d}_H)\l^2(1+(\l^2)^2)}{(\mathrm{d}_G-\mathrm{d}_H)(1+\l^1)^2(1-(\l^2)^2)}\,,\\
&\gamma_2{}^1=\frac{1}{2k}\frac{(c_G(\mathrm{d}_G+\mathrm{d}_H)-2c_H \mathrm{d}_H)(1-\l^1)^2\l^2(1+(\l^2)^2)}{\mathrm{d}_H(1-(\l^2)^2)^3}\,,\\
&\gamma_2{}^2=\frac{c_G}{k}\frac{\l^1(1+3(\l^2)^2)-(\l^2)^2(3+(\l^2)^2)}{(1+\l^1)(1-(\l^2)^2)^2}\,,
\end{split}
\end{equation*}
which transforms as a mixed tensor
under the diffeomorphism endowed by \eqref{sduality}.

\section{Concluding remarks and outlook}

We have been able to compute the anomalous dimensions of a class of integrable models
characterized by two parameters, an integer $k$ and a real parameter $\l$ measuring the deviation from
the CFT WZW model for a semi-simple group $G$.  Our results are leading order in the $\nicefrac1k$ expansion but exact in $\l$.
We have made this possible by a combination of techniques involving the Callan--Symanzik equation,
a non-trivial duality-type symmetry shared by these models, analyticity arguments as well as
the leading order result in the loop expansion in powers of $\lambda$. Our results were shown to be
in perfect agreement with perturbation theory beyond the leading order. Moreover we derived
the all-loop anomalous dimension for the bilinear current operators which deform the exact CFT for the cases: i) isotropic case, ii) the anisotropic
$SU(2)$ case and iii) the two coupling corresponding to the symmetric coset $G/H$ and a subgroup
$H$ splitting of a group $G$.

A natural extension of our work would be to consider cases beyond isotropy, i.e. when the matrix $\l$
is not proportional to the identity. Work in that direction is in progress.
Our methods can be extended for $\l$-deformations based on symmetric coset spaces.
In that case the current wavefunction renormalization is still given by \eqn{zeff}.
Using the $\beta$-function \cite{Itsios:2014lca,Sfetsos:2014jfa, Appadu:2015nfa}
\be
{
\label{RG.coset}
\beta=-\frac{c_G\lambda}{4k}\leqslant0\,,
}
\ee
and plugging into \eqref{anomalous.current} we find that
\be
{
\label{dim.current.coset}
\gamma_J=\frac{c_G\lambda}{2k(1-\lambda^2)}\geqslant0\,,
}
\ee
which is invariant under \eqref{sduality} and also under the $(\lambda,\eta)$-duality it assumes
a real value as easily can be seen by using \eqn{analel}.
It is important to check this result and in that respect the $\b$-function \eqn{RG.coset},
against CFT perturbation theory. The novelty in this case is that we shall need
as basic building blocks correlation function of parafermions because these are driving the perturbation
in these cases \cite{Sfetsos:2013wia}.
We note that unlike the current
case \eqn{dim.current} the correction in \eqn{dim.current.coset} is of ${\cal O}(\l)$.

Our analysis should be extendable to semi-symmetric spaces ($\mathbb{Z}_4$ grading). Indeed,
the computation of the $\beta$-function for the semi-symmetric space
$PSU(2,2|4)/SO(1,4)\times SO(5)$, was performed in
\cite{Appadu:2015nfa}. The result turns out to be proportional to the
quadratic Casimir in the adjoint representation (to leading order in the $1/k$ expansion)
which vanishes for the supergroup $PSU(2,2|4)$. We expect a similar result to hold for the anomalous dimensions
but nevertheless the necessary computation should be performed.
Finally, it would be interesting to identify the operators in the $\eta-$deformed models  of which the anomalous dimension is given by \eqref{eta-anomal}.

\section*{Acknowledgments}

We thank A. Petkou for a useful correspondence.
The research of G. Georgiou was partially
supported by the General Secretariat for Research and Technology of Greece and from
the European Regional Development Fund MIS-448332-ORASY (NSRF 2007-13 ACTION,
KRIPIS). The research of K. Sfetsos is implemented
under the \textsl{ARISTEIA} action (D.654 of GGET) of the \textsl{operational
programme education and lifelong learning} and is co-funded by the
European Social Fund (ESF) and National Resources (2007-2013).
The research of K. Siampos has been supported by the Swiss National Science Foundation and
also acknowledges the \textsl{Germaine de Stael}
France--Swiss bilateral program (project no 32753SG) for financial support.
K. Sfetsos would like to thank
the TH-Unit at CERN for hospitality during the final stages of this project.

\appendix

\section{Abelian part of the metric}

The scope of this appendix is to compute the k-independent (abelian) part of the metrics for the
single and composite operator, given by \eqref{G0} and \eqref{G-iso} respectively. Let us note that
the integrations within this section are performed in the entire complex plane, since there is no need for regularization.

\subsection{Single operator}
\label{singleG}

We firstly write the 2-point function $G_{0|ab}$ as a series in the coupling constant $\lambda$,
i.e. $G_{0|ab}=\sum\limits_{n=0}^\infty \frac{\lambda^{2n}}{\pi^{2n} (2n)!}G_{0|ab}^{(2n)}$.
Notice that the sum in the expansion
for $G_{0|ab}$ runs only in even powers of { the} coupling constant $\lambda$ since any correlator involving odd number of holomorphic or anti-holomorphic currents
vanishes. Subsequently, we derive a recursive relation for the $G_{0|ab}^{(n)}$'s
\begin{equation}
\label{Gn}
\frac{G_{0|ab}^{(n)}}{x_{12}^2}=\int \prod_{i=1}^n \mathrm{d}^2z_i\langle J^{a}(x_1) J^{a_1}(z_1)\cdots J^{a_n}(z_n)  J^{b}(x_2)\rangle
\langle  \bar{J}^{a_1}(\bar{z}_1)\cdots\bar{J}^{a_n}(\bar{z}_n)\rangle\,.
\end{equation}
The next step is to perform the integral with respect to $z_1$. Thus we have to find all the possible ways to contract the two currents
situated at the point $z_1$ with the remaining currents in \eqn{Gn} with the aid of \eqn{OPE}.

There are, actually, two types of contractions.
In the first type we contract both currents at the point $z_1$ only with internal ones while in the second type one of the currents at $z_1$
is contracted with an external one while the other with an internal one.
Thus, in the first type the contraction of $J^{a_1}(z_1)$ is with any of the remaining 'internal' currents sitting at $z_2,z_3,\dots,z_n$.
Apparently there are $n-1$ such possibilities. For convenience let us say that $J^{a_1}(z_1)$ is paired with $J^{a_2}(z_2)$
Then the second current at the point $z_1$, $\bar{J}^{a_1}(\bar{z}_1)$ can be contracted with any
of the $n-2$ currents sitting at $z_3,\dots,z_n$. Note that $\bar{J}^{a_1}(\bar{z}_1)$ can not be contracted with $\bar{J}^{a_2}(\bar{z}_2)$
because this will give rise to bubble diagram and as such has to be omitted. We conclude that one has $(n-1)(n-2)$ such diagrams.
In the second type one may contract $J^{a_1}(z_1)$ with the external holomorphic current situated at $x_1$.
Then the anti-holomorphic current at $\bar{z}_1$ should be paired only with an internal anti-holomorphic one otherwise we shall end up with a disconnected
diagram. Since there are $n-1$ such currents there will be $2(n-1)$ different contractions of this type. The additional factor of 2 comes because
of the two possibilities: $J^{a_1}(z_1)$ can be contracted to $J^{a}(x_1)$ or with $J^{b}(x_2)$.
Using the identity \eqn{id1}
in \eqn{Gn} and performing a second z-integration with the help of the delta-function in \eqn{id3} it is straightforward to obtain
the following recursive relation
\begin{equation}\label{recur}
G_{0|ab}^{(n)}=\pi^2 \big((n-1)(n-2)+ 2(n-1)\big)G_0^{(n-2)}=\pi^2n(n-1) G_{0|ab}^{(n-2)}.
\end{equation}
One can show by induction that the {\it n-th} term in the expansion takes the form
\begin{equation}\label{sol-rec}
G_{0|ab}^{(n)}=\pi^n\,n! G_{0|ab}^{(0)}, \,\,\,\,n=2m,\,\,\,\,m=1,2,3,\dots
\end{equation}
where $G_{0|ab}^{(0)}=\delta_{ab}$ is the free (no interactions) 2-point function.
We are now in position to derive the abelian part of the metric for the single operator
\begin{equation}\label{G0.single}
G_{0|ab}=\delta_{ab}\left(1+\sum_{m=1}^{\infty} \frac{\lambda^{2m}}{\pi^{2m} (2m)!}\pi^{2m} (2m)!\right)=
\frac{\delta_{ab}}{1-\lambda^2}\,,
\end{equation}
which yields  Eq.\eqref{G0}, through $G_{0|ab}:=G_0\,\delta_{ab}$.

\subsection{Composite operator}
\label{compositeG}

In this case $G_0=\sum\limits_{n=0}^\infty \frac{\lambda^{2n}}{\pi^{2n} (2n)!}G_0^{(2n)}$ where:
\begin{equation*}
\frac{G_0^{(n)}}{|x_{12}|^4}=\int \prod_{i=1}^n \mathrm{d}^2z_i\langle J^{a}(x_1) J^{a_1}(z_1)\cdots J^{a_n}(z_n)J^{b}(x_2)  \rangle
\langle \bar{J}^{a}(\bar{x}_1) \bar{J}^{a_1}(\bar{z}_1)\cdots\bar{J}^{a_n}(\bar{z}_n) \bar{J}^{b}(\bar{x}_2)\rangle\,.
\end{equation*}
Working in analogy to the single operator  case, we find the recursive relation:
\begin{equation}
G_0^{(n)}=\pi^2 \big((n-1)(n-2)+ 4(n-1)\big)G^{(n-2)}=\pi^2(n-1)(n+2) G_0^{(n-2)}.
\end{equation}
where the additional factor of 4 comes because
of the four possibilities: $J^{a_1}(z_1)$ can be contracted to $J^{a}(x_1)$ or to $J^{b}(x_2)$ (two possibilities) plus another two possibilities
arising  from the contractions of the anti-holomorphic current at $z_1$ with one of the external anti-holomorphic current at $x_1$ or $x_2$.
One can show by induction that the {\it n-th} term in the expansion takes the form
\begin{equation}\label{sol-rec}
G_0^{(n)}=\frac{\pi^n}{2}(n+2)!! (n-1)!! G_0^{(0)}=\frac{\pi^n}{2}(n+2)n! G_0^{(0)}\, \,\,\,\,n=2m,\,\,\,\,m=1,2,3,\dots\,,
\end{equation}
where $G_0^{(0)}=\dim G$ is the free (no interactions) 2-point function.
We are now in position to derive the abelian part of the metric for the composite operator, i.e. Eq.\eqref{G-iso}:
\begin{equation}\label{G0.composite}
G_0=\dim G \left(1+\sum_{m=1}^{\infty} \frac{\lambda^{2m}}{\pi^{2m} (2m)!}\pi^{2m} \frac{1}{2}(2m+2) (2m)!\right)=
\frac{\dim G}{(1-\lambda^2)^2}\,.
\end{equation}

\end{document}